\documentclass[%
reprint,
superscriptaddress,
showpacs,
 amsmath,amssymb,
 aps,
prb,
]{revtex4-1}

\usepackage{graphicx,subfigure,bbm}
\usepackage{dcolumn}
\usepackage{bm}

\usepackage{amsmath}
\usepackage{amsthm}

\usepackage[english]{babel}
\usepackage[utf8]{inputenc}
\usepackage{graphicx}
\usepackage[colorinlistoftodos]{todonotes}
\usepackage{bbold }
\usepackage{physics}
\usepackage[utf8]{inputenc}
\usepackage{amssymb }
\usepackage{graphicx}
\DeclareGraphicsExtensions{.eps}
\usepackage{epstopdf}
\usepackage{latexsym}
\usepackage{amsfonts}
\usepackage{xcolor}
\usepackage{units}
\usepackage{bm}
\usepackage{verbatim}
\usepackage[colorlinks = true,
            linkcolor = red,
            urlcolor  = blue,
            citecolor = magenta,
            anchorcolor = red]{hyperref}
\usepackage{esint}
\usepackage{soul}
\usepackage{cancel}
\usepackage[normalem]{ulem}
\usepackage{empheq}
\usepackage{dsfont}
\usepackage{soul}

\newcommand{\btheta}{\boldsymbol{\theta}}

\definecolor{AB-color}{RGB}{128,0,0}

\definecolor{GB-color}{RGB}{0,0,128}

\definecolor{FT-color}{RGB}{128,0,128}
\begin{document}


\title{{Nonlocal thermoelectricity in a topological Andreev interferometer}}

\author{Gianmichele Blasi}
\email{gianmichele.blasi@sns.it} 
\affiliation{NEST, Scuola Normale Superiore and Instituto Nanoscienze-CNR, I-56126, Pisa, Italy}
\author{Fabio Taddei}
\affiliation{NEST, Scuola Normale Superiore and Instituto Nanoscienze-CNR, I-56126, Pisa, Italy}
\author{Liliana Arrachea}
\affiliation{International Center for Advanced Studies, ECyT-UNSAM, Campus Miguelete, 25 de Mayo y Francia, 1650 Buenos Aires, Argentina}
\author{Matteo Carrega}
\affiliation{NEST, Scuola Normale Superiore and Instituto Nanoscienze-CNR, I-56126, Pisa, Italy}
\affiliation{SPIN-CNR, Via Dodecaneso 33, 16146 Genova, Italy}
    
\author{Alessandro Braggio}
\email{alessandro.braggio@nano.cnr.it} 
\affiliation{NEST, Scuola Normale Superiore and Instituto Nanoscienze-CNR, I-56126, Pisa, Italy}


\begin{abstract}
We discuss the phase dependent nonlocal thermoelectric effect in a topological Josephson junction in contact with a normal-metal probe.
We show that, due to the helical nature of topological edge states,  nonlocal thermoelectricity is generated by a purely Andreev interferometric mechanism. This {response} can be tuned by imposing a Josephson phase difference, through the application of a dissipationless current between the two superconductors, even without the need of applying an external magnetic field. We discuss in detail the origin of this effect and we provide also a realistic estimation of the nonlocal Seebeck coefficient that results of the order of few $\mu V/K$.
\end{abstract}
\date{\today}

\maketitle
\emph{Introduction}-- 
Prominent topics in hybrid superconducting quantum technologies concern the thermal management \cite{Partanen_2016,Fornieri17,Senior_2020} and thermoelectricity\cite{Mazza_2015,Claughton96,Sanchez18,Kamp19,Hussein19,Kirsanov19,Pershoguba19, Marchegiani20}.
These represent  novel functionalities for quantum sensing\cite{Giazotto15,Hekkila18,Guarcello_2019}, entanglement manipulation~\cite{Blasi_Ent_Manipulation_2019} and thermal engines\cite{Yamamoto15,Mazza14,Marchegiani20b,vischi19}.

Usually, finite thermoelectric response appears in hybrid superconducting systems only when the particle-hole symmetry, encoded in the Bogoliubov-de Gennes (BdG) Hamiltonian, is broken, e.g. by means of ferromagnetic correlations\cite{Machon13,Oazeta14,Kolenda17,shapiro17,keidel20} or by exploiting nonlinearities \cite{Sanchez16,Pershoguba19,Marchegiani20}.
Recently, mechanisms able to generate {\it nonlocal} thermoelectricity have been predicted in Cooper pair splitters\cite{Hussein19,Kirsanov19} and Andreev interferometers~\cite{Virtanen_2004,Titov_2020,Jacquod_2010,Kalenkov_2020}, and experimentally investigated~\cite{Chandrasekhar_1998,Jiang_2005,Petrashov_2003,Tan_2020}. Here, by relying on multiterminal configurations and nonlocal signals, novel mechanisms for topological insulators are opened.
We have demonstrated\cite{Blasi_2020_PRL} that a Josephson junction based on a {two-dimensional} (2D) topological insulator (TI)\cite{Qi2010,ando2013,ti3,ti4,ronetti2017} threaded by a magnetic flux with one edge attached to a normal metallic probe
\cite{bours18}  
presents nonlocal thermoelectricity when a temperature difference is applied between the two superconducting leads.
The responsible mechanism is the so-called Doppler shift induced by the magnetic flux in the junction, 
which has an effect akin to a Zeeman splitting in the two spin-polarized members of the Kramer pair of the 2D TI \cite{teoj6}. 

In this work, we show that a phase bias alone in a topological Josephson junction is sufficient to establish finite nonlocal thermoelectricity. 
This is very appealing since, differently from the mechanism {of Ref. \onlinecite{Blasi_2020_PRL}, which requires the presence of a magnetic field, the present one relies on a purely interferometric effect. Its origin is the helical property of edge states and the superconducting phase difference.{ Importantly}, such thermoelectric response disappears when both edges are connected to the probe or when the TI is replaced by normal channels. Hence, it constitutes a peculiar feature of the helical nature of 2D TI. We argue that, with state-of-the-art technologies it leads to a nonlocal Seebek coefficient of the order of few $\mu V/K$.

\emph{Model}--We consider the topological Josephson junction (TJJ) depicted in Fig.~\ref{scheme_with_temperature_voltage}, which consists of two superconducting electrodes placed on top of a 2D TI at a distance $L$.
The two electrodes induce superconducting correlations on the edge states via proximity effect~\cite{teoj6,ther1}.
The width of the TI strip is assumed to be large enough such that upper and lower edges are decoupled, and we focus only on the upper edge.
The system is described by the following BdG Hamiltonian 
\begin{equation}
\label{My_Hamiltonian}
 {\cal H}=\mqty(H(x) & i\sigma_y\Delta (x) \\ -i\sigma_y\Delta (x)^*  &  -H(x)^*) ,
\end{equation}
expressed in the Nambu basis $(\psi_{\uparrow},\psi_{\downarrow},\psi_{\uparrow}^{*},\psi_{\downarrow}^{*})^T$ with spin $\uparrow$ and $\downarrow$ collinear with natural spin-quantization axis of the TI edge along $z$-direction, where $H(x)=v_F\left(-i\hbar\partial_x \right)\sigma_z-\mu\sigma_0$ with $-H(x)^*$ being its time-reversal partner. 
The Fermi velocity is $v_{\rm F}$, $\mu$ is the chemical potential and $\sigma_i$ are the Pauli matrices. We consider rigid boundary conditions with order parameter $\Delta(x) = \Delta_0\left[\Theta(-x)e^{i\phi_{\rm S_L}} + \Theta(x - L)e^{i\phi_{\rm S_R}}\right]$, where $\Theta(x)$ is the step function, $\Delta_0$ is the proximity induced gap and $\phi\equiv\phi_{\rm S_R}-\phi_{\rm S_L}$ is the gauge invariant Josephson phase difference between the two superconductors.
A normal-metal probe N -- such as a STM tip~\cite{das2011,liu2015,hus2017,voigtlander2018} -- is directly contacted to the upper edge on the point $x_0$ (see Fig.~$\ref{scheme_with_temperature_voltage}$)
and modeled by an energy- and spin- independent transmission amplitude $t$.

\begin{figure}[h!!]
\centering
    \includegraphics[width=.5\textwidth]{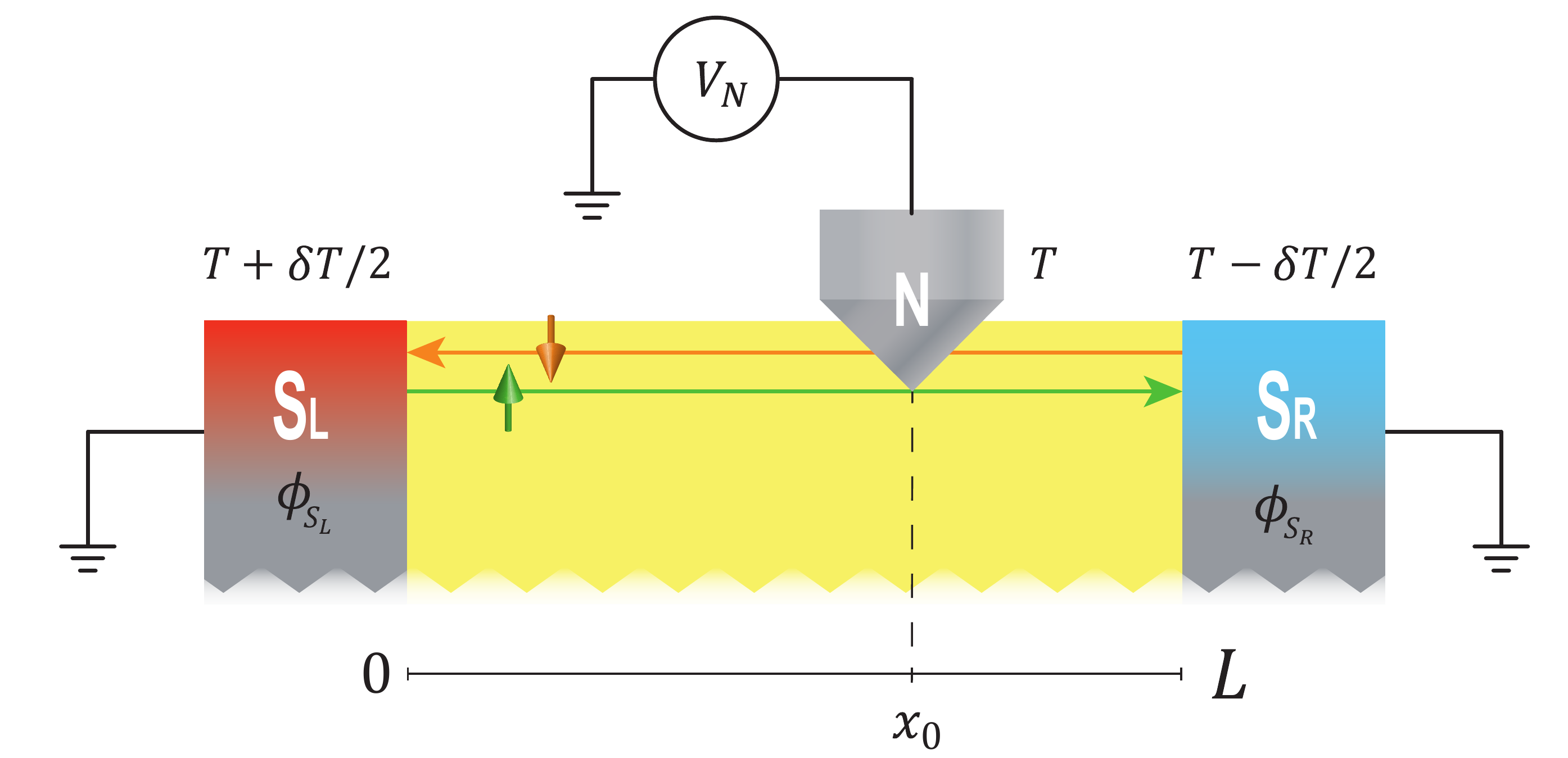}
    \caption{A helical Kramers pair of edge states of the quantum spin Hall effect is contacted by two superconductors at different temperatures $T_{S_L}=T+\delta T/2$ and $T_{S_R}=T-\delta T/2$, and a phase difference $\phi\equiv \phi_{S_R}-\phi_{S_L}$. A bias voltage $V_N$ is applied to the normal-metal probe at temperature $T_N=T$ and coupled to the edge at the point $0\leq x_0\leq L$, with $L$ the length of the junction. 
} 
    \label{scheme_with_temperature_voltage} 
\end{figure}

\emph{Charge and Heat Currents}-- In the setup depicted in Fig.~\ref{scheme_with_temperature_voltage}, a voltage bias $V_N$ is applied between the probe $N$ and the superconducting electrodes (grounded) and a thermal bias $\delta T=T_{S_L}-T_{S_R}$ is imposed between the left and right superconductors, while the temperature of the probe is $T_{N}=(T_{S_L}+T_{S_R})/2=T$ . 
In this configuration the relevant responses are the charge current $J_N^c$ flowing in the probe and the heat current $J_{\rm S_L}^h$ flowing in the left superconductor~\cite{linear_no_heat}.
Within the scattering approach~\cite{Lambert98,Blanter00} 
they read
\begin{equation}
\label{current_Lambert_general}
\begin{pmatrix}
J_N^c   \\   
J_{\rm{S_L}}^h
\end{pmatrix}=\frac{2}{h}\!\!\sum_{j,\alpha,\beta}\!\int_{0}^{\infty}\!\!\!\!\!\! d\epsilon 
\begin{pmatrix}
 \alpha e \left(f_{N}^{\alpha}(\epsilon)-f_j^{\beta}(\epsilon)\right)P_{N,j}^{\alpha,\beta}(\epsilon,\phi)\\
\epsilon\left(f_{S_L}^{\alpha}(\epsilon)-f_j^{\beta}(\epsilon)\right)P_{S_L,j}^{\alpha,\beta}(\epsilon,\phi)
\end{pmatrix}
\end{equation}
where $\alpha,\beta=+$ stand for quasi-particle (QP), $\alpha,\beta=-$ for quasi-hole (QH), and with $j$ running over leads indices ($S_L$, $S_R$ and $N$). In Eq.~($\ref{current_Lambert_general}$) we assumed the chemical potentials of the grounded superconductors as reference for the energies. 
The currents depend on the generalized Fermi distributions $f^{\alpha}_j(\epsilon)=\{e^{(\epsilon-\alpha eV_j )/k_BT_j}+1\}^{-1}$, where $T_j$ and $V_j$ are respectively the temperature and the voltage at the lead $j$. Notice that when $ V_j=0$ (as for the grounded superconductors $V_{S_L}=V_{S_R}=0$), $f^-_j(\epsilon)=f^+_j(\epsilon)$.
The scattering coefficients $P_{i,j}^{\alpha,\beta}(\epsilon,\phi)$, with $i,j=N,S_L,S_R$, represent the reflection ($i=j$) or transmission ($i\neq j$) probabilities of a quasi-particle of type $\beta$ in lead $j$ to a quasi-particle of type $\alpha$ in lead $i$. As a consequence of the helical nature of the edge states and the spin independence of the probe, it turns out that these probabilities do not depend on the position $x_0$ of the probe (hence all the results discussed hereafter do not depend on the probe position).  

Furthermore, 
for $\delta T,V_N\rightarrow0$ it is possible to write the currents of Eq.~($\ref{current_Lambert_general}$) in linear order with respect to these parameters in the following form\cite{Benenti17,Mazza14,Roura18,Hussein19,Sanchez18,Kirsanov19}
\begin{align}
\label{Osanger}
J_N^c&=L_{11} (V_N/T) + L_{12}(\delta T/T^2)\nonumber\\
J_{S_L}^h&=L_{21} (V_N/T) + L_{22}(\delta T/T^2) .
\end{align}
Interestingly, although the configuration contains three terminals, the driving affinities are only two namely $V_N/T$ and $\delta T/T^2$. Hence, the Onsager matrix, with entries $L_{ij}$, is effectively $2\times2$~\cite{Mazza14,linear_no_heat,Benenti17,Sanchez15,Roura18,Mani2018}. In this effective formulation one should remind that $L_{12}$ and $L_{21}$
are \emph{nonlocal} thermoelectrical coefficients.

\emph{Symmetries}-- It is known~\cite{Lambert98,Jacquod12} that the scattering coefficients $P_{ij}^{\alpha\beta}$ satisfy relations due to microreversibility $P_{ij}^{\alpha\beta}(\epsilon,\phi)=P_{ji}^{\beta\alpha}(\epsilon,-\phi)$, particle-hole symmetry $P_{ij}^{\alpha\beta}(\epsilon,\phi)=P_{ij}^{-\alpha-\beta}(-\epsilon,\phi)$ and unitarity $\sum_{\alpha i}P_{ij}^{\alpha\beta}(\epsilon)=N_j^{\beta}(\epsilon)$, $\sum_{\beta j}P_{ij}^{\alpha\beta}(\epsilon)=N_i^{\alpha}(\epsilon)$.
Here $N_i^{\alpha}(\epsilon)$ is the number of open channels for $\alpha$-type quasiparticles at energy $\epsilon$ in lead $i$. Nevertheless, the helical nature of the edge, the symmetry between left/right superconducting gaps (which are equal in case of linear thermal bias regime) and the fact that the coupling to the probe is independent of energy and spin result in additional symmetries. In particular the reflection coefficients at the probe $N$ satisfy the relation $P_{\rm NN}^{\alpha\beta}(\epsilon,\phi)=P_{\rm NN}^{-\alpha-\beta}(\epsilon,\phi)$ between QP and QH states. Further, there are also peculiar \emph{nonlocal} symmetries of the scattering coefficients between the probe and the left/right superconductors, namely $P_{\rm NS_{L/R}}^{\alpha\beta}(\epsilon,\phi)=P_{\rm NS_{R/L}}^{-\alpha-\beta}(\epsilon,\phi)$ and  $P_{\rm NS_{L/R}}^{\alpha\beta}(\epsilon,\phi)=P_{\rm NS_{L/R}}^{-\alpha-\beta}(\epsilon,-\phi)$~\cite{footnote_symmetries}.

\emph{Nonlocal thermoelectric response}
-- By exploiting the aforementioned symmetry relations, one can write the charge current at the probe $J_N^c$ in the following form:
\begin{equation}
\label{J00}
J_N^{c}=\frac{2}{h}\!\!\int_{0}^{\infty}\!\!\!\!\!\!d\epsilon~\Big\{ F_N^-(\epsilon) A(\epsilon,\phi)- F_S^-(\epsilon)\left[Q(\epsilon,\phi)- Q(\epsilon,-\phi)\right]\Big\}
\end{equation}
where in the first term we recognize the Fermi function differences for normal probe
$F_N^-\equiv f_N^+-f_N^-$ weighted with a scattering coefficient \begin{equation}
\label{A}
A (\epsilon,\phi)=e\left(N_N^+- P^{++}_{NN}+P^{+-}_{NN}\right)=e\left(N_N^--P^{--}_{NN}+P^{-+}_{NN}\right)
\end{equation}
that represents the electronic charge transferred from the probe $N$ into the edge, being $P^{\pm\pm}_{NN}$ normal reflections and $P^{\pm\mp}_{NN}$ the Andreev ones.
The second term instead contains the Fermi function differences between the two superconductors
$F_S^{-}\equiv f_{\rm{S_L}}^{\pm}-f_{\rm{S_R}}^{\mp}$ which are non-zero when a thermal bias $\delta T\neq0$ is applied between the superconductors.
The function $F_S^-$ is weighted with the odd parity component, with respect to $\phi$, of  the function
\begin{equation}
\label{Q}
Q (\epsilon,\phi)=e\left(P^{++}_{NS_L}- P^{-+}_{NS_L}\right)=-e\left(P^{+-}_{NS_R}- P^{--}_{NS_R}\right)
\end{equation}
A visualization of the meaning of the quantity $Q$  is given in Fig.~$\ref{paths1}$ where we sketch the resonant processes where a QP or QH is injected from right or left superconductors and is transferred after multiple resonant Andreev processes to the probe as an electron (solid) or a hole (dashed). In particular $Q$ represents the net electronic charge transferred into the probe $N$ when a QP is injected from $S_L$ (see Fig.~$\ref{paths1}$ $(a)$). The symmetries show that a QH injected from the right superconductor $S_R$ brings exactly the same amount of charge, with opposite sign [second identity of Eq.~(\ref{Q})] as represented in Fig.~$\ref{paths1}$ $(b)$. Alongside these processes (represented  in Figs.~$\ref{paths1}$ $(a)$-$(b)$), there are also dual processes, depicted in Fig.~$\ref{paths1}$ $(c)$-$(d)$, which correspond to the same amount of transferred charge given in Eq.~($\ref{Q}$) obtained  by exchanging the side of injection (i.~e.~$S_L\rightleftarrows S_R$) and inverting the sign of $\phi\rightarrow-\phi$.

\begin{figure}[h!!]
\centering
    \includegraphics[width=.5\textwidth]{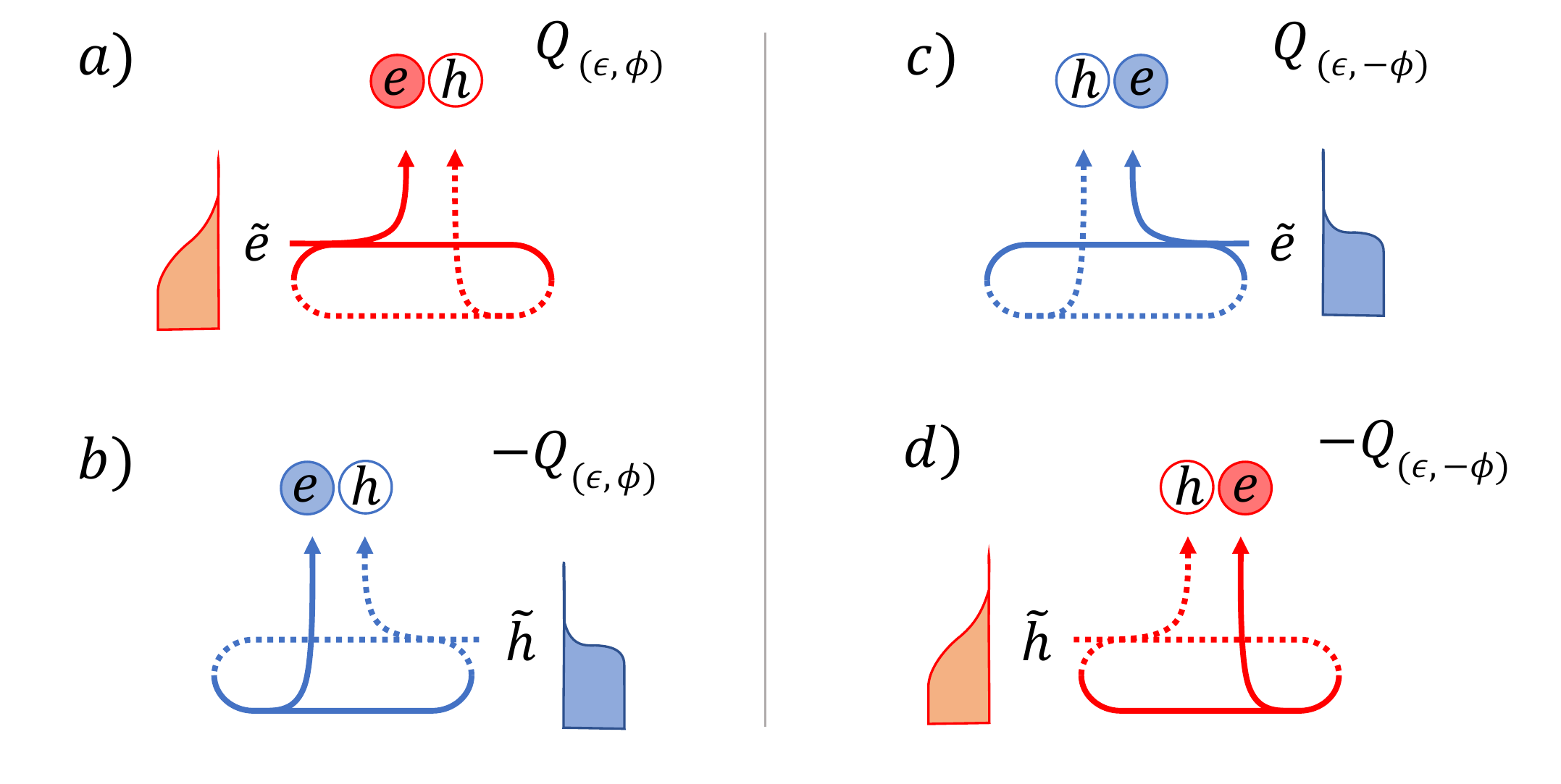}
    \caption{Resonant processes describing the transfer of the charge $Q$ from the superconducting leads $S_L,S_R$ into the probe $N$. $\tilde{e}$,$\tilde{h}$ label respectively QP and QH at the superconducting leads. Solid and dashed lines correspond to the trajectories traveled by electrons and holes respectively.  
Red (blue) correspond to processes originated at the hot (cold) lead $S_L$ ($S_R$) whose Fermi distribution $f_{S_L}=f_{S_L}^\pm$ ($f_{S_R}=f_{S_R}^\pm$) is sketched on the side.
In $(a)$-$(b)$ are depicted the processes of QP and QH injected from $S_L$ and $S_R$ respectively, and corresponding to a transfer of the opposite amount of charge $Q(\epsilon,\phi)$~$(a)$ and $-Q(\epsilon,\phi)$~$(b)$. In $(c)$-$(d)$ are depicted the dual processes obtained by inverting the lead of injection ($S_L\rightleftarrows S_R$) and the sign of $\phi\rightarrow -\phi$.   
} 
    \label{paths1} 
\end{figure}

We now discuss the physical consequence of the result reported in Eq.~($\ref{J00}$). 
When $V_N=0$ there is no contribution from the Fermi functions of the normal probe (i.~e.~$F_N^-=0$) since $f_N^+(\epsilon)=f_N^-(\epsilon)$.
Since $T_N$ does not enter these expressions, the possibility of inducing {\em local} thermoelectricity by means of a thermal gradient between the TI and the probe is ruled out.
This is particularly important at experimental level since the temperature of the probe does not need to be controlled during the measurement of nonlocal thermoelectricity. 

The only thermoelectric response in the probe is the $\emph{nonlocal}$ one when a thermal bias between the two superconductors $\delta T$ is applied, i.e. $F_S^-=f_{\rm{S_L}}^{\pm}(\epsilon)-f_{\rm{S_R}}^{\mp}(\epsilon)\neq0$.
This nonlocal thermoelectric response (see Eq.~($\ref{J00}$)) is determined by the integral over the energies of the odd parity component in $\phi$ of the function $Q(\epsilon,\phi)$, i.e. $Q(\epsilon,\phi)- Q(\epsilon,-\phi)$. If $\phi=0$, one cannot have nonlocal thermoelectricity.  The physical reason of this result comes from the exact cancellation of the contributions of the processes represented in Fig.~$\ref{paths1}$: in particular $(a)$ cancels with $(d)$ and (b) with (c).

\emph{Phase dependent thermoelectricity}--
Here we concentrate on the action of the Josephson phase bias $\phi$ showing that
is responsible for the generation of nonlocal thermoeletricity in the probe due to a peculiar Andreev interferometric effect associated to the helical nature of the edge, as pictorially sketched in Fig.~$\ref{paths1}$.
This can be rationalized looking at the analytical expressions of the quantities $A$ and $Q$ of Eqs.~($\ref{A}$) and ($\ref{Q}$):
\begin{align}
\label{A_anal}
A(\epsilon,\phi)&=\sum_{\sigma=\pm}\frac{2e\abs{t}^4\cdot \Theta(\Delta-\epsilon)}{1+\abs{r}^4+2\abs{r}^2\cos{(2\pi\frac{L \epsilon}{\xi \Delta}+\sigma\phi+2\arcsin{(\frac{\epsilon}{\Delta})})}}\nonumber\\
&+\sum_{\sigma=\pm}\frac{e(g(\epsilon)+1)(g(\epsilon)-\abs{r}^2)\abs{t}^2\cdot \Theta(\epsilon-\Delta)}{g(\epsilon)^2+\abs{r}^4-2g(\epsilon)\abs{r}^2\cos{(2\pi\frac{L \epsilon}{\xi \Delta}+\sigma\phi)}}
\end{align}
\begin{align}
\label{Q_anal}
Q(\epsilon,\phi)&=\frac{e(g(\epsilon)-1)(g(\epsilon)-\abs{r}^2)\abs{t}^2\cdot \Theta(\epsilon-\Delta)}{g(\epsilon)^2+\abs{r}^4-2g(\epsilon)\abs{r}^2\cos{(2\pi\frac{L \epsilon}{\xi \Delta}-\phi)}}
\end{align}
where $g(\epsilon)=\left(\epsilon/\Delta+\sqrt{\epsilon^2/\Delta^2-1}\right)^2$, $\abs{r}^2=1-\abs{t}^2$ and $\xi=\hbar v_F/\pi\Delta$ is the superconducting coherence length.  
Notice that Eq.~($\ref{A_anal}$) consists of two parts each related to the sub-gap (first line) and supra-gap (second line) processes, while Eq.~($\ref{Q_anal}$) contains only the supra-gap contribution.
In particular, from Eq.~($\ref{Q_anal}$), it emerges that $Q(\epsilon,\phi)$ has no definite symmetry in $\phi$ for $L \neq 0$ so that one would expect a finite nonlocal thermoelectric response.

\begin{figure}[h!!]
\centering
    \includegraphics[width=.5\textwidth]{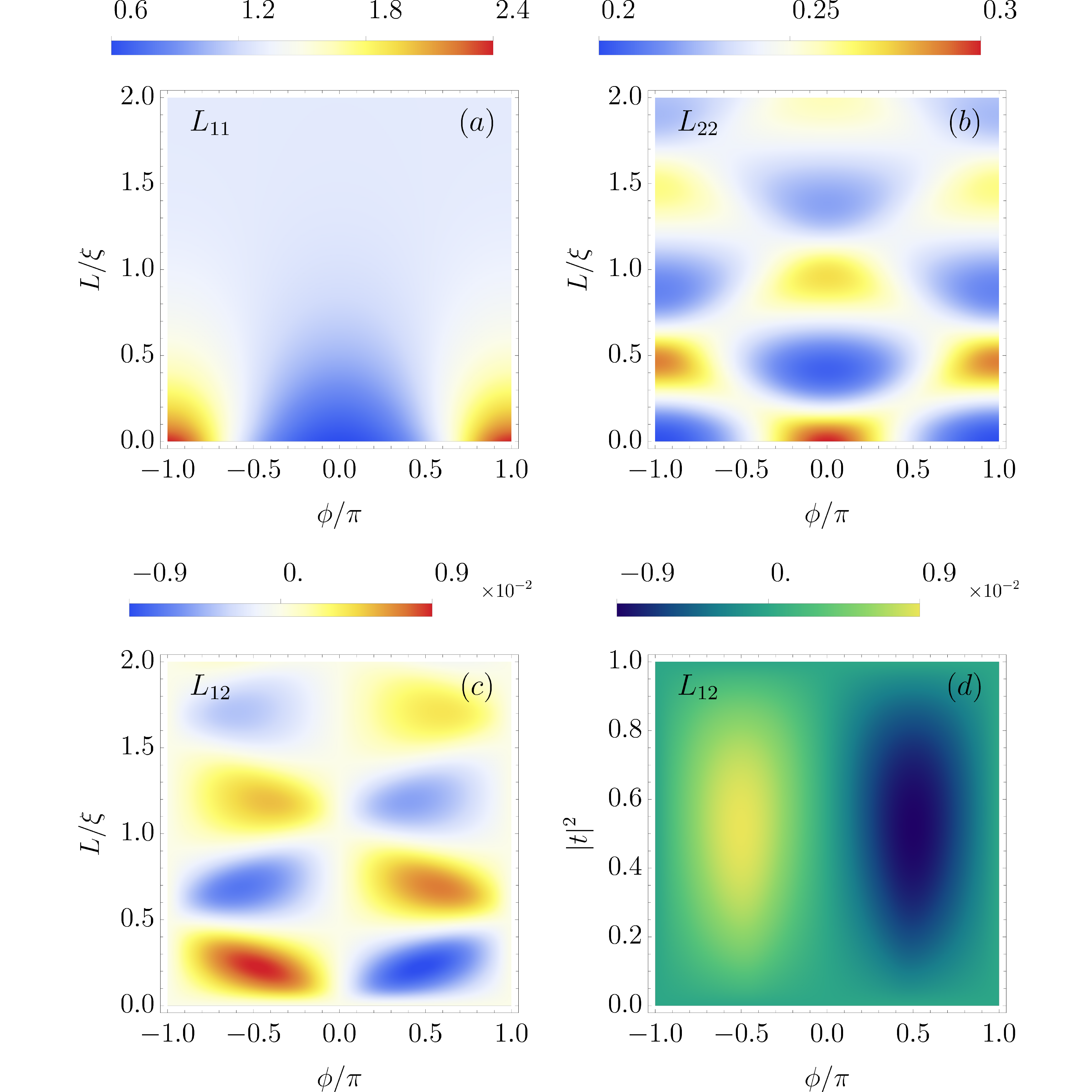}
    \caption{Phase dependence  of the Onsager coefficients. $L_{11}$ (a), $L_{22}$ (b) and $L_{12}=-L_{21}$ (c)
	as functions of $\phi/\pi$ and the junction length  $L/\xi$ for $|t|^2=0.5$ . (d) $L_{12}$ as a function of  $\phi/\pi$ and coupling parameter $|t|^2$ with the junction length $L/\xi=0.25$ (for which is maximal) . Such quantities are taken at $T/T_C=0.4$ and normalized as follows: $L_{11}/(G_0 T)$, $L_{22}/(G_T T^2)$ and $L_{12}/(\sqrt{G_0 G_T T^3})$, with $G_0=2e^2/h$ and $G_T=(\pi^2/3h)k_B^2 T$ being respectively the electrical conductance quantum and the thermal conductance quantum.} 
    \label{Onsager_without_DS} 
\end{figure}

The interferential nature of the phenomena can be better enlightened by investigating the behaviour of the Onsager coefficients as a function of the junction length $L$. In Fig.~$\ref{Onsager_without_DS}$ the Onsager coefficients are plotted as functions of $\phi/\pi$ and the length measured as $L/\xi$. In Figs.~\ref{Onsager_without_DS}$(a)$,$(b)$ and $(c)$ we plot respectively, the local Onsager coefficients $L_{11}$, $L_{22}$ and the nonlocal thermoelectrical coefficient $L_{12}$ setting the strength of the coupling with the probe at an intermediate value $|t|^2=0.5$ and the temperature fixed at $T/T_c=0.4$ (the highest temperature at which the induced gap of the right and left superconductors remain constant and equal to $\Delta_0$). Notice that, by exploiting the aforementioned symmetries of the scattering coefficients, it can be shown that the off-diagonal nonlocal coefficients satisfy the relation $L_{12}(\phi)=L_{21}(-\phi)=-L_{21}(\phi)$, similarly to the case discussed in Ref.~\onlinecite{Blasi_2020_PRL}.

We observe that $L_{11}$ (which is proportional to the conductance at the probe), is an even function~\cite{Jacquod12} of $\phi$ and, for small length $L\ll\xi$, presents a minimum for $\phi\approx 0$ and a maximum at $\phi\approx \pm\pi$. Increasing the length $L$, the conductance become featureless and flat due to an effective averaging between the (increasing) number of available states involved in the transport. 
More interesting, instead, is the behavior of $L_{22}$ and $L_{12}$ which present a periodicity of one coherence length $\xi$ as functions of the length of the junction~\cite{ther1}.
This periodicity is determined by the oscillatory change of available states at energies $\epsilon\gtrsim\Delta$, which dominate 
the spectral contribution to the transport window, oscillating between a maximum to a minimum when the junction length changes by one $\xi$ length. This effect is not present in $L_{11}$ since it is mostly determined by subgap states given by the Andreev contributions. Remarkably, the thermal conductance ($\propto L_{22}$) crucially differs from the nonlocal thermoelectric coefficient ($\propto L_{12}$) since the first is even with the phase bias $\phi$ while the latter is odd~\cite{Jacquod12,Richter_2011} (see Eq.~($\ref{J00}$)). 
The different symmetry in $\phi$ is due to the fact that QPs and QHs contribute with the same sign to the heat transport but with opposite sign to the thermoelectric current. 

In Fig.~$\ref{Onsager_without_DS}$ $(d)$ we show how $L_{12}$ changes with the coupling parameter $|t|^2$, keeping the length of the junction fixed to $L/\xi=0.25$ (for which it is maximal, see Fig.~$\ref{Onsager_without_DS}$ $(c)$). It emerges that  the absolute value of the nonlocal Onsager coefficient $L_{12}$ reaches its maximum for an intermediate value of the coupling parameter (i.~e.~$\abs{t}^2\approx 0.5$), while is zero either when $\abs{t}^2\approx0$ (when the probe is decoupled) or $\abs{t}^2\approx1$ (when the two superconductors are mutually decoupled, and individually well coupled to the $N$ probe). 

We stress that the appearance of this nonlocal linear thermoelectric effect is a unique feature of our hybrid topological Josephson junction when the probe is contacted with just one helical edge of the TI.
Such thermoelectric response, instead, disappears when both edges are connected to the probe or when the TI is replaced by normal spinfull channels.
More precisely, we verified that in the case of a standard $S$-$N$-$S$ junction in contact with a normal-metal probe, $L_{12}=L_{21}=0$. 

\begin{figure}[h!]
\centering
    \includegraphics[width=.5\textwidth]{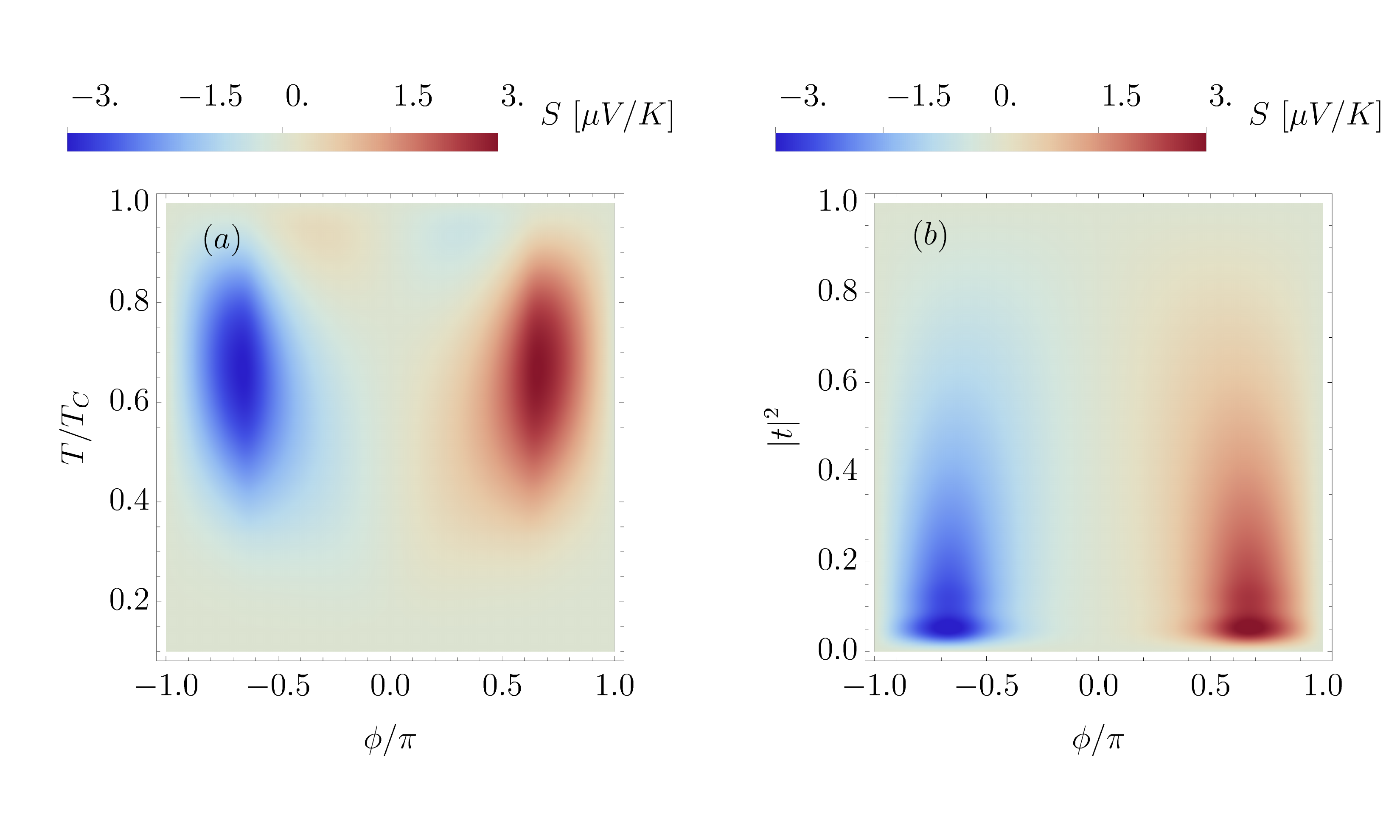}
    \caption{$(a)$ Nonlocal Seebeck coefficient as function of $\phi/\pi$ versus $T/T_C$ for $\abs{t}^2=10^{-2}$. $(b)$ Nonlocal Seebeck coefficient as function of $\phi/\pi$ versus the probe coupling $|t|^2$ for $T/T_C=0.7$. Both $(a)$-$(b)$ have been obtained for the same length $L/\xi=0.25$.} 
    \label{Seebeck} 
\end{figure}

As a final remark, it is important to give a realistic estimation of the strength of the thermoelectrical effect we are discussing. 
In this regard, we compute the nonlocal Seebeck coefficient $S=(1/T)L_{12}/L_{11}$~\cite{Benenti17} as a function of $\phi$ (see Fig.~$\ref{Seebeck}$). In order to make realistic predictions in a wide temperature range, here we also include the temperature dependence of the gap order parameter~\cite{delta_footnote}.
Fig.~$\ref{Seebeck}$(a) shows that the nonlocal Seebeck coefficient grows with the operating temperature and reaches a maximum of $3 \mu V/K$ roughly at $T/T_C\approx 0.7$ for $\phi/\pi\approx\pm 0.6$. At higher temperatures the gap closes reducing the nonlocal thermoelectricity, hence confirming the fundamental role of the superconducting state. 
Fig.~$\ref{Seebeck}$(b) (obtained for $T/T_C=0.7$) shows how the nonlocal Seebeck effect scales with the probe coupling $|t|^2$ observing that  small value of the tunnelling with the probe returns highest values. 
Notably these values of the phase-dependent nonlocal Seebeck coefficient are roughly $\approx 6\%$ of the values determined by the Doppler shift mechanism proposed in Ref.~\onlinecite{Blasi_2020_PRL}. The advantage in the present case, is that there is  no need to apply any magnetic field to measure it since it is enough to impose a dissipationless current between the two superconductors to induce the phase bias $\phi$.
 This experimental protocol seems quite attractive due to its simplicity and the absence of any spurious Nernst effect~\cite{Nernst_1886,Behnia_2016,Zuev_Chang_Kim_2009,Zhu_2010}.

In Figs.~$\ref{Seebeck}$ we considered the length of the junction $L/\xi=0.25$. This situation is reasonable assuming a STM tip with state-of-the-art size of 100 nm and a coherence length $\xi$ in the proximized TI of the order of 600 nm~\cite{Hart14,Bocquillon18}. Further, this choice of the length assures that the transport along the edge state is ballistic~\cite{Groenendijk18} at the operating temperatures for our setup, typically of a few K.

\emph{Conclusions}--We have investigated a phase-dependent nonlocal thermoelectricity in a topological Josephson junction coupled to a probe. 
We have shown that an Andreev interferometric mechanism  affects QPs and QHs differently resulting in a nonlocal thermoelectric response. We have discussed the dependence of this mechanism over the junction length $L$ and the coupling with the probe $|t|^2$. We have estimated, with realistic parameters, a nonlocal Seebeck coefficient of few $\mu V/K$.
Nevertheless, we underline that the provided estimations are quite conservative since the critical temperature $T_C$ of the induced proximized gap is given by the critical temperature of the parent superconductors which is usually much higher further increasing the nonlocal Seebeck coefficient which is proportional to operating temperature.
As in the case of Ref. \onlinecite{Blasi_2020_PRL}, this thermoelectric effect is a consequence of the helical nature of the edge states. Therefore, it can be used as an evidence of the existence of these states in TI systems.
Although the present Seebeck coefficient is one order of magnitude smaller than the one predicted in  Ref. \onlinecite{Blasi_2020_PRL}, it has the advantage of taking place 
 in absence of any magnetic field, hence, under much simpler experimental conditions. 

\begin{acknowledgments}
 We acknowledge support from CNR-CONICET cooperation program ``Energy conversion in quantum nanoscale hybrid devices''.  We are sponsored by PIP-RD 20141216-4905 
of CONICET,   PICT-2017-2726 and PICT-2018-04536 from Argentina, as well as the Alexander von Humboldt Foundation, Germany and the ICTP-Trieste through a Simons associateship (LA). A.B. and F.T acknowledge SNS-WIS joint lab QUANTRA. M. C. is supported by the Quant-Era project ``Supertop''. A. B. acknowledge the Royal Society through the International Exchanges between the UK and Italy (Grant No. IEC R2 192166).
\end{acknowledgments}

%
%


\begin{thebibliography}{9}
\bibitem{footnote_symmetries}{All these symmetry relations are still valid in presence of a magnetic flux $\Phi$, under the substitution $\phi\rightarrow\btheta\equiv(\phi,\Phi)$.}
\bibitem{delta_footnote}{We took $\Delta(T)=\Delta_{0}\tanh{\left(1.74\sqrt{T_C/T-1}\right)}$, an approximation good at 2\% with respect to the self-consistent BCS result~\cite{Kamp19,Tinkham}.}
\end{thebibliography}

\end{document}